\documentclass[twocolumn,superscriptaddress,prl,floatfix]{revtex4}
\usepackage[]{graphicx}
\usepackage[usenames,dvipsnames]{color}
\usepackage{soul}
\usepackage{bm}
\usepackage{amsmath}
\usepackage{dcolumn}
\usepackage{color}
\usepackage{bm}
\usepackage{youngtab}

\newcommand{\nuQ}{\nu_\mathrm{Q}}
\newcommand{\gamman}{\gamma_\mathrm{n}}
\newcommand{\Hext}{H_\mathrm{ext}}

\newcommand{\Vx}{\emph{V}_\mathrm{xx}}
\newcommand{\Vy}{\emph{V}_\mathrm{yy}}
\newcommand{\Vz}{\emph{V}_\mathrm{zz}}

\newcommand{\cubbio}{BaTi$_2$As$_2$O}

\begin{document}

\title{Revealing the hidden order in {\cubbio} via nuclear magnetic resonance}

\author{D. W. Song}
\affiliation{Hefei National Laboratory for Physical Sciences at the Microscale, University of Science and Technology of China, Hefei, Anhui 230026, China}

\author{J. Li}
\affiliation{Hefei National Laboratory for Physical Sciences at the Microscale, University of Science and Technology of China, Hefei, Anhui 230026, China}

\author{D. Zhao}
\affiliation{Department of Physics, University of Science and Technology of China, Hefei, Anhui 230026, China}

\author{L. K. Ma}
\affiliation{Hefei National Laboratory for Physical Sciences at the Microscale, University of Science and Technology of China, Hefei, Anhui 230026, China}

\author{L. X. Zheng}
\affiliation{Hefei National Laboratory for Physical Sciences at the Microscale, University of Science and Technology of China, Hefei, Anhui 230026, China}

\author{S. J. Li}
\affiliation{Hefei National Laboratory for Physical Sciences at the Microscale, University of Science and Technology of China, Hefei, Anhui 230026, China}

\author{L. P. Nie}
\affiliation{Hefei National Laboratory for Physical Sciences at the Microscale, University of Science and Technology of China, Hefei, Anhui 230026, China}

\author{X. G. Luo}
\affiliation{Hefei National Laboratory for Physical Sciences at the Microscale, University of Science and Technology of China, Hefei, Anhui 230026, China}
\affiliation{Department of Physics, University of Science and Technology of China, Hefei, Anhui 230026, China}
\affiliation{Key Laboratory of Strongly-coupled Quantum Matter Physics, University of Science and Technology of China, Chinese Academy of Sciences, Hefei 230026, China}
\affiliation{Collaborative Innovation Center of Advanced Microstructures, Nanjing 210093, China}

\author{Z. P. Yin}
\email{yinzhiping@bnu.edu.cn}
\affiliation{Department of Physics and Center for Advanced Quantum Studies, Beijing Normal University, Beijing 100875, China}

\author{T. Wu}
\email{wutao@ustc.edu.cn}
\affiliation{Hefei National Laboratory for Physical Sciences at the Microscale, University of Science and Technology of China, Hefei, Anhui 230026, China}
\affiliation{Key Laboratory of Strongly-coupled Quantum Matter Physics, University of Science and Technology of China, Chinese Academy of Sciences, Hefei 230026, China}
\affiliation{Collaborative Innovation Center of Advanced Microstructures, Nanjing 210093, China}
\author{X. H. Chen}
\affiliation{Hefei National Laboratory for Physical Sciences at the Microscale, University of Science and Technology of China, Hefei, Anhui 230026, China}
\affiliation{Department of Physics, University of Science and Technology of China, Hefei, Anhui 230026, China}
\affiliation{Key Laboratory of Strongly-coupled Quantum Matter Physics, University of Science and Technology of China, Chinese Academy of Sciences, Hefei 230026, China}
\affiliation{Collaborative Innovation Center of Advanced Microstructures, Nanjing 210093, China}

\date{\today}
\begin{abstract}

In low-dimensional metallic systems, lattice distortion is usually coupled to a density-wave-like electronic instability due to Fermi surface nesting and strong electron-phonon coupling(M. D. Johannes and I. I. Mazin, Phys. Rev. B 77, 165135(2008)). However, the ordering of other electronic degrees of freedom can also occur simultaneously with the lattice distortion thus challenges the aforementioned prevailing scenario. Recently, a hidden electronic reconstruction beyond FSN was revealed in a layered metallic compound {\cubbio} below the structural transition temperature $\emph{T$_S$}\sim200K$. The nature of this hidden electronic instability is under strong debate. Here, by measuring the local orbital polarization through $^{75}$As nuclear magnetic resonance experiment, we observe a $p$-$d$ bond order between Ti and As atoms in {\cubbio} single crystal. Below \emph{T$_S$}, the bond order breaks both rotational and translational symmetry of the lattice. Meanwhile, the spin-lattice relaxation measurement indicates a substantial loss of density of states and an enhanced spin fluctuation in the bond-order state. Further first-principles calculations suggest that the mechanism of the bond order is due to the coupling of lattice and nematic instabilities. Our results strongly support a bond-order driven electronic reconstruction in {\cubbio} and shed light on the mechanism of superconductivity in this family.

\end{abstract}


\maketitle


Superconductivity in titanium-based oxypnictides BaTi$_2$Pn$_2$O (Pn= As, Sb, Bi) and their isovalent as well as aliovalent analogys: BaTi$_2$(Sb$_{1-x}$Bi$_x$)$_2$O and Ba$_{1-x}$A$_x$Ti$_2$Sb$_2$O (A= Na, K, Ru), was recently discovered\cite{9,10,11,12,13,14,15}. Due to their quasi two-dimensional (2D) layered structure of Ti$_2$O-plane resembling both cuprates and iron-based superconductors, much attention have been attracted to these materials. In particular,  they show a remarkable anomaly in both temperature-dependent charge transport and magnetic susceptibility, which are ascribed to electronic instabilities like charge- or spin-density wave (CDW/SDW)\cite{3,10,16,17,18}. By substitution of As with isovalent Sb or Bi, the density wave transition is gradually suppressed and finally leads to superconductivity. This suggests a similar competition between superconductivity and the density wave order as in cuprates and iron-based superconductors\cite{10}. Therefore, identifying the exact nature of the possible density wave order in {\cubbio} is not only crucial for understanding itself, but also of great importance to understand the superconductivity in this family.

To uncover the electronic ordering in titanium-based oxypnictides BaTi$_2$Pn$_2$O (Pn= As, Sb) and Na$_2$Ti$_2$Pn$_2$O (Pn= As, Sb), much efforts have been made in previous studies. Early electronic structure calculations predicted that the Fermi surfaces of this family are quite well nested and therefore susceptible to either CDW or SDW instabilities\cite{18,19,20,21,22}. Furthermore, recent calculations on the phonon dispersion also supported lattice instability and Fermi surface nesting driven CDW in these materials\cite{6,23}. Experimentally, most of experiments in this family have failed to give direct evidence for CDW or SDW order\cite{5,24,25}. Moreover, instead of superlattice distortion, only a rotational symmetry is broken below the structural transition temperature in early neutron scattering experiment of {\cubbio}\cite{5}. Recently, superlattice distortion has been successfully observed in Ba$_{1-x}$Na$_x$Ti$_2$Sb$_2$O and Na$_2$Ti$_2$Pn$_2$O (Pn = As, Sb), which strongly supports the charge density wave picture\cite{26,27}. However, in contrast to conventional Fermi surface nesting scenario, the observed partial gap in angle resolved photoemission spectrum (ARPES) experiment opens on the so-called ``Fermi patches" rather than the Fermi surface, and accompanies an anomalous spectral weight redistribution within a large energy scale\cite{4,27,28}. Theoretically, various local electronic orderings have been proposed beyond charge density wave picture, including charge/orbital ordering within one unit cell of Ti$_2$O square\cite{5,7} and spin-driven bond ordering\cite{8}. In these cases, structural transition is driven by electronic instability similar to that in FeSe\cite{29}. Up to date, direct experimental confirmation for the electronic order is still missing in this family. Especially, how to verify the possible ordering phenomenon in orbital degree of freedom is even challenging. Here, we use nuclear magnetic resonance (NMR) technique to investigate the local orbital polarization at As sites in {\cubbio} single crystals. For NMR technique, when the nuclear spin number (\emph{I$_S$}) is equal to 1/2, the Knight shift extracted from NMR central transition line mainly reflects the magnetic hyperfine interaction between nuclear and electrons, which is related to local spin susceptibility ($\chi_s$), such as $^{77}$Se (\emph{I$_S$} $=$ 1/2) in FeSe\cite{30,31}. However, when \emph{I$_S$} is larger than 1/2, the Knight shift extracted from NMR central transition line also have an additional contribution from the second-order effect of the nuclear quadrupole interaction (NQI). Considering the NQI, the NMR frequency of the central transition line is expressed as\cite{32}:
\begin{equation}
\label{equ:nuQ}
    \nu = \gamman H(K+1) +\frac{f(\nuQ,\eta,\theta,\phi)}{\gamman H(K+1)},
\end{equation}

The first term is from the magnetic contribution, while the second term is from the quadrupole contribution. More detailed information about Eq(1) is in the caption of Fig.1. In this work, through a comprehensive study on the quadrupole contribution of $^{75}$As (\emph{I$_S$} $=$ 3/2) NMR spectra, we successfully obtained the information on orbital polarization and revealed a bond order (BO) in {\cubbio} single crystal.

\begin{figure*}[!t]
 \centering
 \includegraphics[width=\textwidth,angle=0,clip=true]{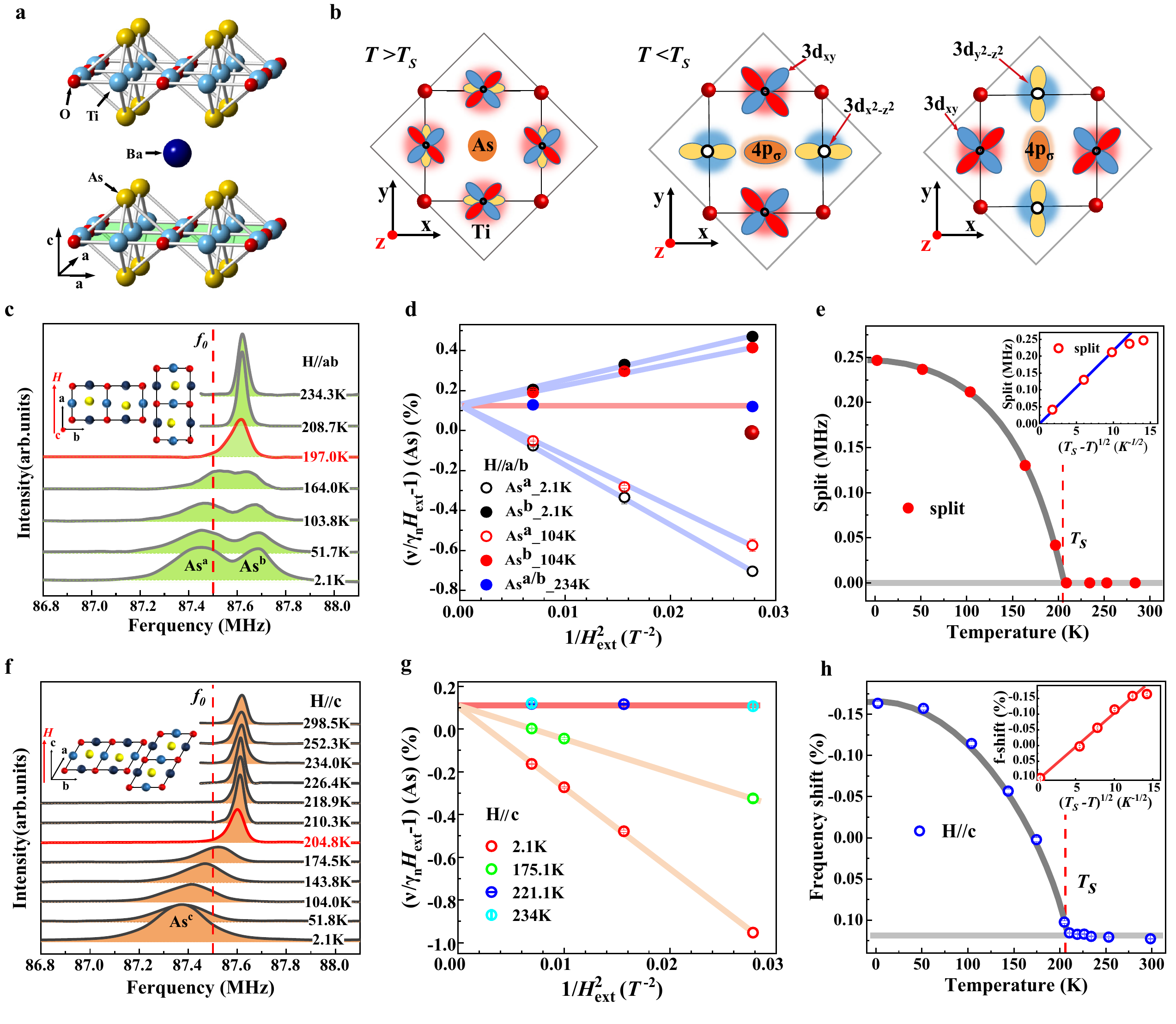}
 \caption{\textbf{Evidence for the tetragonal-to-orthorhombic structural transition and  4\emph{p$_\sigma$} orbital polarization in {\cubbio}}.
 (\textbf{a}) Crystallographic structure of {\cubbio} with [Ti$_2$As$_2$O]$^{2-}$ and Ba$^{2+}$ layers stacked alternatively along $c$-axis. (\textbf{b}) The possible orbital polarization configuration in each structural domains above and below structural transition temperatue \emph{T$_S$}. Below \emph{T$_S$}, the polarization of 4\emph{p$_\sigma$} orbitals at As sites and 3$d$ orbitals with lowest energies at Ti sites\cite{5,7} breaks rotational symmetry. (\textbf{c}) and (\textbf{f}) Temperature dependent $^{75}$As NMR central lines measured at $\Hext\parallel[100]$ ($\Hext\parallel a/b$) and $\Hext\parallel[001]$ ($\Hext\parallel c$). Below \emph{T$_S$}, there is a substantial quadrupole contribution on the central transition line. Then, the total NMR frequency is described by the Eq(1). \emph{K} represents the Knight shift tensor due to the magnetic hyperfine interaction. $\nuQ$ is the quadrupole frequency, defined as $\nuQ =\frac{eQ\Vz}{\hbar}$. $\eta$ is the asymmetry parameter, defined as $\eta=|\frac{\Vx-\Vy}{\Vz}|$ , where $|\Vx|\leq |\Vy|\leq |\Vz|$ are the principal components of the electric field gradient (EFG) at nuclei site. $\theta$ as well as $\phi$ are the polar angles of the applied magnetic field ($\Hext$) with respect to the principal axis frame which is assumed to be the same for both the Knight shift and EFG tensors. (\textbf{d}) and (\textbf{g}) The field dependent $(\frac{\nu}{\gamman\Hext}-1)$ of As$^{a}$/As$^{b}$ and As$^{c}$ at different temperatures. The bold lines are guiding for eyes. (\textbf{e}) The temperature-dependent splitting of NMR central line with $\Hext\parallel a/b$, and (h) the negative frequency-shift of NMR central line with $\Hext\parallel c$. Both of them can be taken as an order parameters, showing typical $\sqrt{T_s-T}$ behaviour of a second-order phase transition (see the inset).}
 \label{fig1}
 \end{figure*}

As shown in Fig.1c and Fig.1f, the temperature-dependent $^{75}$As NMR central lines are measured in {\cubbio} single crystal under two different configurations with $\Hext\parallel[100]$ (Fig.1c) and $\Hext\parallel[001]$ (Fig.1f). Above \emph{T$_S$}, the central transition lines are narrow and symmetric for both configurations, suggesting a high quality of our sample. As temperature drops below \emph{T$_S$}, the central transition line with $\Hext\parallel[100]$ splits into two lines, assigned as As$^{a}$ ($\Hext\parallel a$) and  As$^{b}$ ($\Hext\parallel b$). Such splitting is ascribed to the tetragonal-to-orthorhombic structural transition at \emph{T$_S$}, because the twinning of the orthorhombic domains makes each domain point either along the direction of $\Hext$ or orthogonal to $\Hext$, as sketched in Fig.1c. As shown in Fig.1e, if we take the splitting as an order parameter, the temperature-dependent splitting is proportional to $\sqrt{T_S-T}$ which is consistent with a second-order phase transition in Landau theory. Furthermore, the structural orthorhombicity revealed by previous neutron diffraction is perfectly scaled with the observed NMR line splitting, suggesting that both of them are related to the same Landau order parameter. When $\Hext\parallel[001]$ or [110], all orthorhombic domains are equivalent; hence, we only observe a single central transition line in orthorhombic phase below \emph{T$_S$} (see Fig.1f and Fig.S3 in Supplemental Material).

Similar phenomenon is also observed for the tetragonal-to-orthorhombic structural transition in $^{75}$As NMR study of LaFeAsO\cite{33}, in which the splitting of the central transition line is dominated by the second order effects of the NQI. Based on Eq(1), while the magnetic contribution on NMR frequency is proportional to $\Hext$, the quadrupole contribution on NMR frequency should be proportional to $\frac{1}{\Hext}$. As shown in Fig.1d, the field-dependent effective Knight shift $(\frac{\nu}{\gamman\Hext}-1)$ confirms that the splitting of the central transition line with $\Hext\parallel[100]$ is also dominated by the second order effects of the NQI. The extrapolated magnetic Knight shift at $\Hext \to \infty$ only has a tiny contribution on the total splitting of the central transition line (more details see Fig.S4 in Supplementary information) and is not our focus in the present study. Furthermore, although there is no splitting in the central transition line with $\Hext\parallel[001]$, a significant negative frequency shift appears below \emph{T$_S$} which is also due to the second order effects of the NQI (Fig.1g). As shown in Fig.1h, such negative frequency shift follows the same temperature dependence as that of the splitting with $\Hext\parallel[100]$. Usually, when the magnetic field is applied along the principal axis of the electric field gradient (EFG) tensor, the quadrupole contribution on the NMR frequency of the central transition line disappears unless asymmetry parameter $\eta$ is nonzero. In a previous NQR study, $\eta \neq 0$ due to the tetragonal-to-orthorhombic structural transition has already been observed below \emph{T$_S$} on BaTi$_2$Sb$_2$O\cite{24}. However, the negative slopes for both As$^{a}$ and As$^{c}$ in Fig.1d and Fig.1g suggest that, besides $\eta \neq 0$, the principal axis of the EFG tensor must be rotated away from [001] direction which is the principal axis of the EFG tensor in the tetragonal phase above \emph{T$_S$} (see Fig.S7 in Supplemental Material). All these results suggest that there is a remarkable change in EFG tensor after structural transition at \emph{T$_S$}.

In principle, the origin of EFG tensor can be divided into two parts\cite{34}: one part comes from the neighboring ions as point charges, which is defined as lattice contribution; the other part comes from the on-site electrons of partially filled orbitals, which is defined as valence contribution. Based on the point charge model, we have studied the lattice contribution to the total EFG tensor (see Fig.S9 in Supplemental Material). We found that both of the tetragonal-to-orthorhombic structural transition and the possible charge order at Ti sites\cite{5} have negligible effect on the EFG tensor through lattice contribution. Therefore, the remarkable change in EFG tensor below \emph{T$_S$} ($\nuQ\approx 5.34$ MHz, see the next section for details) should be ascribed to the valence contribution at $^{75}$As sites. As mentioned above, the valence contribution is related to the partially filled orbitals. When the partially filled orbitals are spherically symmetric or nonpolarized in total, the valence contribution on EFG tensor is zero. In order to explain the significant enhancement of EFG tensor due to valence contribution below \emph{T$_S$}, a strong polarization of the partially filled orbitals is needed. Considering the 4\emph{p$_\sigma$} ($\sigma$ $=$ $x$, $y$, $z$) orbitals as the partially filled orbitals at $^{75}$As sites, our results strongly support that there is a significant orbital polarization for 4\emph{p$_\sigma$} orbitals at $^{75}$As sites below \emph{T$_S$}, which is probably due to the $p$-$d$ bond ordering between Ti and As atoms. In the next section, through a comprehensive study on the angular dependent NMR spectra, the $p$-$d$ bond ordering between Ti and As atoms is further confirmed.

\begin{figure*}[!t]
 \centering
 \includegraphics[width=\textwidth,angle=0,clip=true]{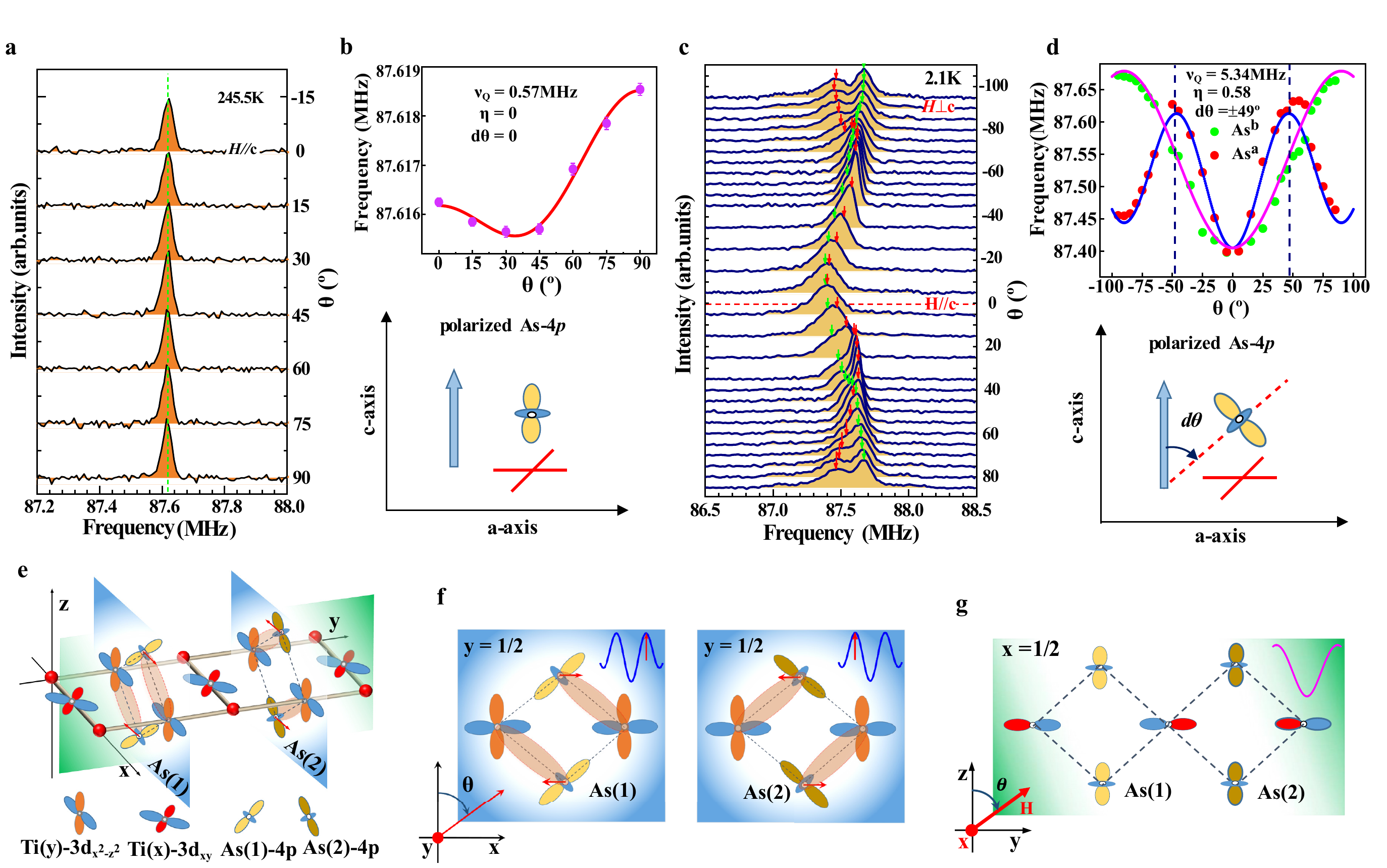}
 \caption{
 \textbf{Evidence for $Ti$-$As$ bond ordering in {\cubbio}}. (\textbf{a}) and (\textbf{c}) Azimuthal-angle dependence of $^{75}$As NMR central lines at tetragonal and orthorhombic phase, respectively. The $\theta$ is defined as the angle between the applied magnetic field $\Hext$ and the crystallographic $c$ axis, where $\theta = 0$ is corresponding to $\Hext\parallel c$. The red and green arrows are the guides to the eyes to show the evolution of central frequency positions of As$^{a}$ and As$^{b}$ domains with respect to $\theta$. (\textbf{b}) and (\textbf{d}) The angular dependence of the central frequency position extracted from Fig.2\textbf{a} and Fig.2\textbf{c}, respectively, which can be perfectly fitted by Eq(2) (solid line). The corresponding polarization of 4$p$ orbital of As is illustrated in the lower panel. (\textbf{e}) Schematic of the $p$-$d$ bond ordering (BO) between Ti and As atoms with the possible orbital order on Ti sites. (\textbf{f}) and (\textbf{g}) Illustration of the angular dependent experiment for As$^{a}$ and As$^{b}$ domains based on the established BO model in (\textbf{e}), the top-left insert solid curves (colorful) are the corresponding angular dependence of the central frequency positions from (\textbf{d}). From the point of symmetry, the alternative arrangement of As(1) and As(2) makes the BO model here is not uniquely determined, and leads to other possible situations, see Supplemental Material.}
 \label{fig2}
 \end{figure*}

As shown in Fig.2, we further performed an angular dependent measurement on $^{75}$As NMR central transition line at 245.5K (tetragonal phase, Fig.2a) and 2.1K (orthorhombic phase, Fig.2c), respectively. Above \emph{T$_S$}, the angular dependent central transition line shows a very tiny angular dependence in tetragonal phase, as shown in Fig.2a, which can be fitted well by Eq(1) (see Fig.2b) and gives $\nuQ \approx 0.57MHz$, $\eta = 0$ and $d\theta = 0$. Here, d$\theta$ is the angle between the principal axis of the EFG tensor and the crystallographic $c$ axis. Based on the discussions on EFG tensor above, the small value of $\nuQ$ suggests a possible weak polarization of 4\emph{p$_\sigma$} orbitals above \emph{T$_S$}. However, below \emph{T$_S$}, the central transition lines for both As$^{a}$ and As$^{b}$ domains show a strong angular dependence and are highly symmetrical with respect to $\theta = 0$ ($\Hext\parallel c$), see Fig.2c.
The angular dependent central position of As$^{a}$ (red dots) and As$^{b}$ (green dots) domains are shown in Fig.2d. For As$^{a}$ domain, there are two symmetrical maximums located at  \(\theta \approx 49^{\circ}\) and  \(\theta \approx -49^{\circ}\) respectively, which are ascribed to the principal axis of the EFG tensor (more details see Supplemental Material). According to the definition of EFG tensor\cite{32}, this result suggests that there are two different $^{75}$As sites in As$^{a}$ domain, assigned as As(1) and As(2). Since the principal axis of EFG tensor for these two $^{75}$As sites are symmetrical respect to $\theta = 0$ and the angle between them is close to 90$^{\circ}$, their angular dependent central positions have a similar behavior. Therefore, considering a realistic line broadening effect, they always overlap together and can not be directly distinguished from the spectra (see Fig.S8 in Supplementary information). Another supporting evidence for two inequivalent $^{75}$As sites with different principal axis of EFG tensor is from the symmetrical linewidth respect to $\theta = 0$, which is not consistent with only one $^{75}$As site (see Fig.S8 in Supplemental Material). For As$^{b}$ domain, there should be also two different $^{75}$As sites as As$^{a}$ domain. Since the principal axis of EFG tenor for these two different $^{75}$As sites are out of the rotational plane and symmetrical respect to the rotational plane, these two different $^{75}$As sites have exactly the same angular dependence and are always overlapped.
Furthermore, we quantitatively analyze the angular dependent central positions of both As$^{a}$ and As$^{b}$ domain by taking the average of the NMR frequency for both As(1) and As(2) sites with modified Eq(1) as below:
\begin{equation}
\label{equ:nu}
   \nu=\gamman H(K+1) +\frac{f(\nuQ,\eta,\theta,d\theta,\phi)+f(\nuQ,\eta,\theta,-d\theta,\phi)}{\gamman H(K+1)},
\end{equation}
where the second term is the average of the quadrupole contribution for both As(1) and As(2) sites. Both of the central frequency positions for As$^{a}$ and As$^{b}$ domains can be fitted well by Eq(2)(see Fig.2d) and gives the same fitting parameters $\nuQ \approx 5.34$MHZ, $d\theta = 49^{\circ}$ and $\eta =0.58$. Compared to the tetragonal phase, $\nuQ$ increases by about one order of magnitude from 0.57MHz to 5.34MHz, supporting a significant change of EFG tensor at $^{75}$As sites. As we discussed above, such a drastic change of EFG tensor is due to the polarization of 4\emph{p$_\sigma$} orbitals at $^{75}$As sites, suggesting an enhanced covalent bond between Ti and As atoms. Another important information from the angular dependent measurement is $d\theta = 49^{\circ}$, which is almost the same as the angle of 48.74$^{\circ}$ between $Ti$-$As$ bond direction and the $c$ axis. This unambiguously proves that the principal axis of the EFG tensor is rotated from $c$-axis to one of the $Ti$-$As$ bond direction, suggesting a $Ti$-$As$ BO.
As shown in Fig.2e, we proposed a BO model with rotational and translational symmetries breaking. Below \emph{T$_S$}, the BO between Ti and As atoms shifts the As atoms from the cental axis of the Ti$_2$O square towards one of the Ti sites, assigned as Ti(y). Such local structural distortion breaks the rotational symmetry coinciding with the tetragonal-to-orthorhombic structural transition in this system. Moreover, since there are two different $^{75}$As sites in each orthorhombic domain, the long-ranged BO should also break the translational symmetry of the lattice as shown in Fig.2e. This is beyond the previous neutron scattering result\cite{5}. Such translational symmetry breaking is also confirmed by very recent Raman and resonant X-ray scattering experiments\cite{35}, which also support the existence of two different As sites in each orthorhombic domain. Finally, based on the BO model, the physical explanation for the angular dependent central position of As$^{a}$ and As$^{b}$ domains can be explicitly understood as shown in Fig.2f and Fig.2g. We should emphasis that the BO model proposed in Fig.2e is not the only choice and other BO models with rotational and translational symmetry breaking could also interpret our results (see Supplemental Material). The present NMR results definitely prove a $p$-$d$ BO between Ti and As atoms in {\cubbio} below the structural transition temperature, which breaks both rotational and translational symmetries of the lattice.

\begin{figure}[!t]
 \centering
 \includegraphics[width=\columnwidth,angle=0,clip=true]{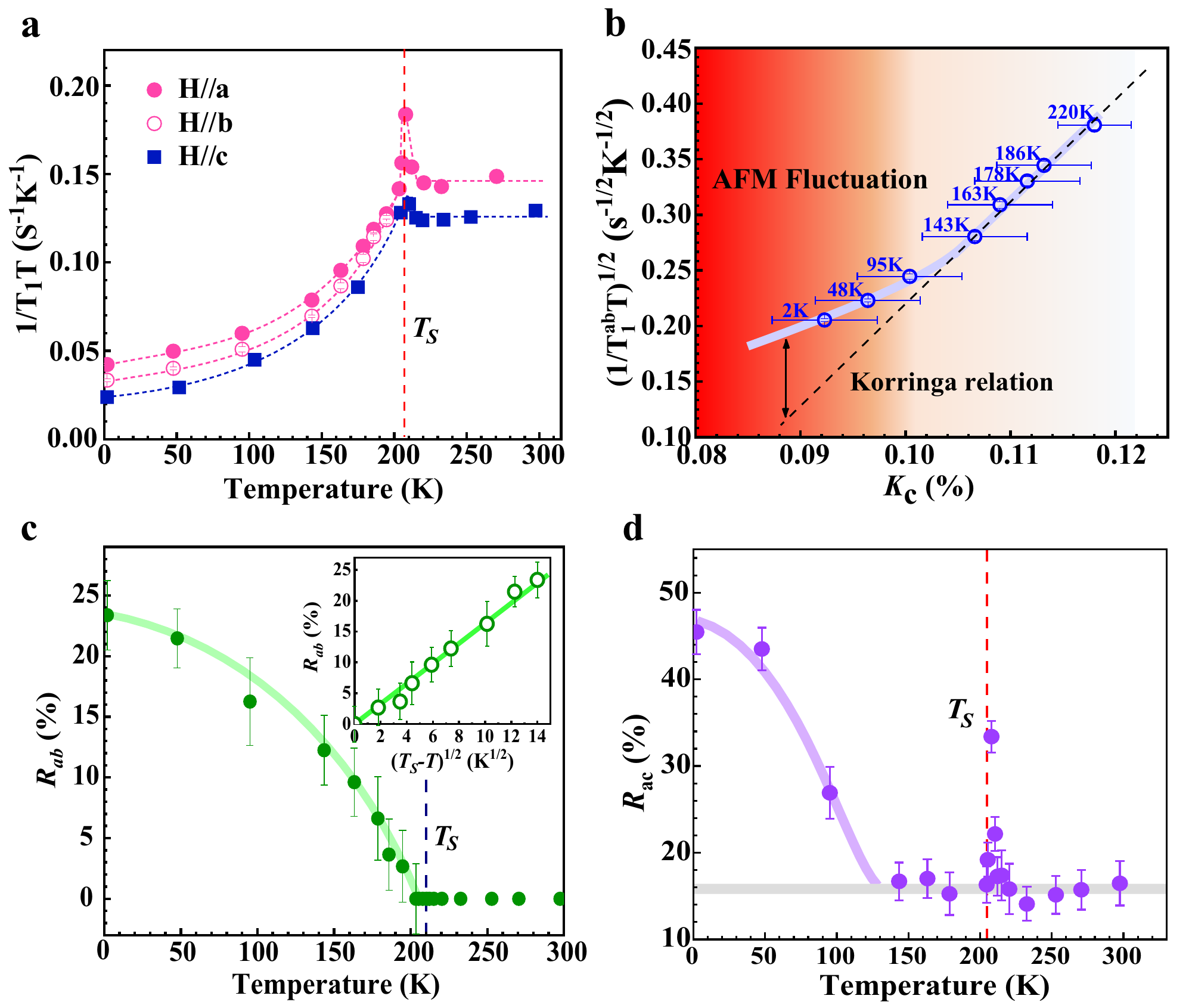}
 \caption{
 \textbf{Evidence for gap opening and enhanced spin fluctuation in bond ordering state.} (\textbf{a}) The temperature-dependent $(T_{1}T)^{-1}$ for $\Hext\parallel c$ and $\Hext\parallel ab$. (\textbf{b}) $(1/(T_{1}T))^{-1/2}$ with $\Hext\parallel ab$ vs $K_{c}$ plot. The dash line shows the Korringa relation. The deviation from Korringa relation is clearly observed below ~100K. Moreover, the upturn behavior is consistence with an enhanced antiferromagnetic(AFM) fluctuation as similar as that in FeSe\cite{31}. (\textbf{c}) The defined out-of-plane anisotropy $R_{ac}$ and (d) the in-plane anisotropy $R_{ab}$ of spin-lattice relaxation rate deduced from the data in (a). The out-of-plane anisotropy ratio is defined as:
 $R_{ac}=2\times\frac{\frac{1}{2}\times[(T_{1}T)_a^{-1}+(T_{1}T)_b^{-1}]-(T_{1}T)_c^{-1}}{\frac{1}{2}\times[(T_{1}T)_a^{-1}+(T_{1}T)_b^{-1}]+(T_{1}T)_c^{-1}}\times100\%$ and the in-plane anisotropy ratio is defined as: $R_{ab}=2\times\frac{(T_{1}T)_a^{-1}-(T_{1}T)_b^{-1}}{(T_{1}T)_a^{-1}+(T_{1}T)_b^{-1}}\times100\%$. The inset in (c) shows typical $\sqrt{T_s-T}$ behaviour of a second-order Landau phase transition. The sharp peak around \emph{T$_S$} in (\textbf{d}) indicates that the critical spin fluctuation is also $ab$-plane polarized AFM fluctuation as that below ~100K.}
 \label{fig3}
 \end{figure}

Fig.3a shows the temperature-dependent spin-lattice relaxation rate divided by temperature 1/$(T_{1}T)$. Above \emph{T$_S$}, the temperature-dependent 1/$(T_{1}T)$ shows a very weak temperature dependence. Upon cooling, a sharp peak due to critical fluctuation shows up in temperature-dependent 1/$(T_{1}T)$ around \emph{T$_S$}.Below \emph{T$_S$}, 1/$(T_{1}T)$ drops quickly and finally becomes saturated at low temperature. In general, the total relaxation rate could be divided as $1/(T_{1}T)=1/(T_{1}T)_{FL}+1/(T_{1}T)_{SF}$. $1/(T_{1}T)_{FL}$ is the contribution from a Fermi liquid and $1/(T_{1}T)_{SF}$ is due to additional spin fluctuations. For a Fermi liquid,$1/(T_{1}T)_{FL}\propto K_{s}^{2}\propto N^{2}(E_{F})$, which is called Korringa relation. As shown in Fig.3b, the Korringa relation is satisfied above $\sim$100K. This suggests that the initial loss of 1/$(T_{1}T)$ below \emph{T$_S$} is due to the suppression of density of states (DOS) at Fermi level, which is consistent with a partial energy gap in the previous ARPES experiment\cite{4}. Below $\sim$100 K, the $(1/T_{1}T)^{1/2}$ vs $K_{c}$ plot clearly deviates from the Korringa relation, suggesting the rising of antiferromagnetic fluctuations.
In order to further clarify the antiferromagnetic fluctuations, we plot the temperature-dependent anisotropy of 1/$(T_{1}T)$. As shown in Fig.3c and 3d, the temperature dependencies for $R_{ac}$ and $R_{ab}$ are quite different. Below \emph{T$_S$}, the value of $R_{ab}$ is immediately increased and shows a similar $\sqrt{T_S-T}$ behavior as that of the NMR splitting, which is ascribed to the change of hyperfine coupling tensor due to the BO. However, the value of $R_{ac}$ only changes below $\sim$100K instead of \emph{T$_S$}, which can not be explained by the change of hyperfine coupling tensor alone. According to the symmetry analysis of hyperfine coupling tensor and the remarkable anisotropic critical spin fluctuation around \emph{T$_S$}, the enhancement of the $R_{ac}$ below $\sim$100K is ascribed to $ab$-plane polarized antiferromagnetic fluctuations with a possible $Q_{AF}$ centered at $(\pi,\pi)$ (see Supplemental Material for details), which suggests an enhanced antiferromagnetic fluctuations at low temperature. Theoretically, spin-driven nematic orbital/bond order has already been proposed for BaTi$_2$Pn$_2$O (Pn = As, Sb) family\cite{7,8}, which should boost the spin fluctuations below nematic transition temperature as similar as that in Fe-based superconductor\cite{30,31,33}. Therefore, our spin-lattice relaxation result suggests an important role of the spin-driven nematic instability on the bond order in {\cubbio}.

\begin{figure}[!t]
 \centering
 \includegraphics[width=1.05\columnwidth,angle=0,clip=true]{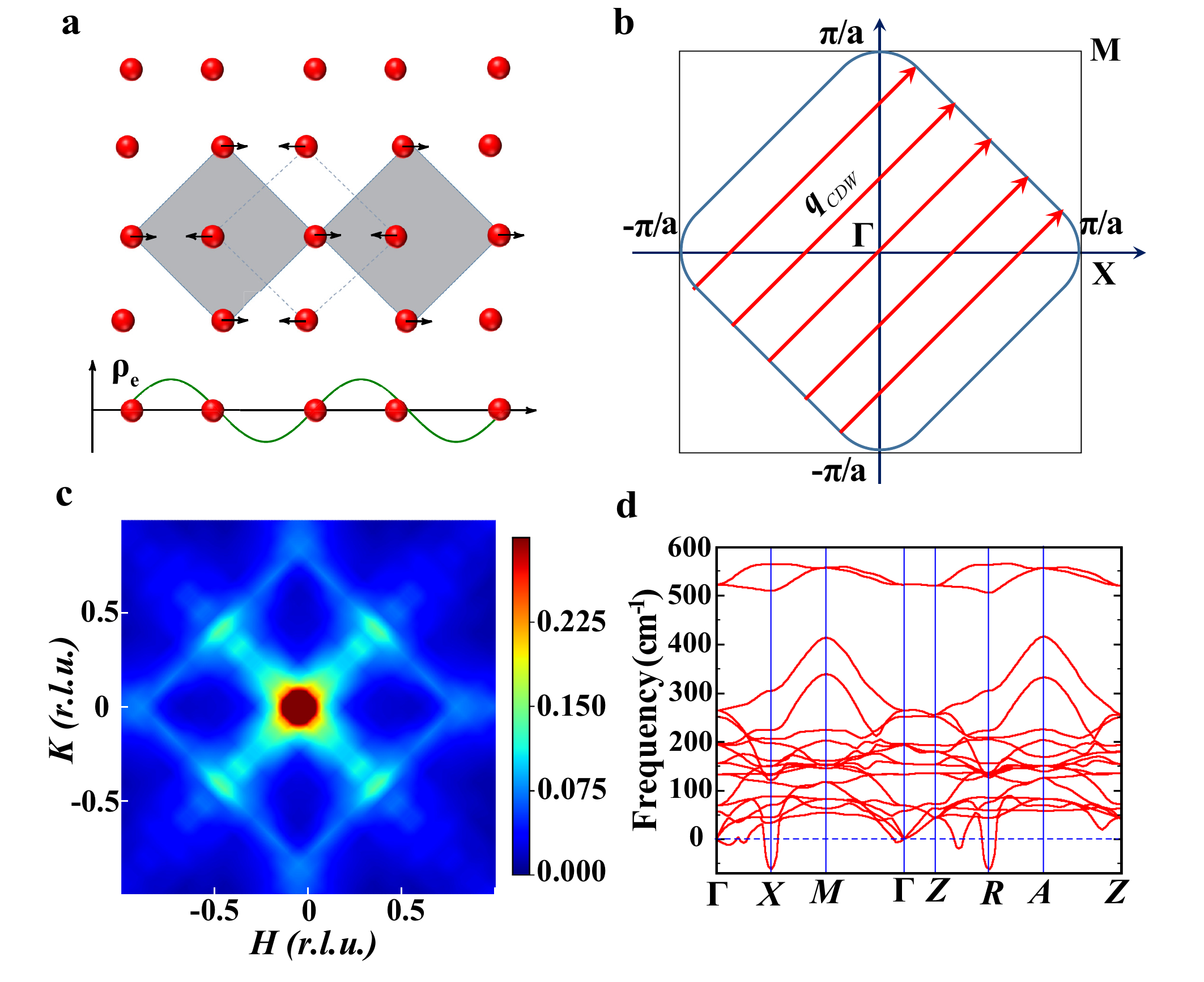}
 \caption{
 \textbf{Conventional CDW picture and theoretical investigation on the lattice and nematic instabilities of {\cubbio}}. Schematic of traditional CDW in a quasi-2D structure accompanied with (\textbf{a}) supperlattice distortion and (\textbf{b}) Fermi surface nesting. (\textbf{c}) The calculated 2D momentum-dependent spin excitations at $\omega=5$ meV. (\textbf{d}) The calculated phonon dispersions of {\cubbio}. Here, all the calculations were performed using tetragonal $P$4/mmm symmetry.}
 \label{fig4}
 \end{figure}

To further explore the mechanism of the bond order, we carry out first-principles calculations on {\cubbio} with tetragonal crystal structure. As shown in Fig.4d, the lattice dynamics calculation indicates a negative (imaginary) frequency around $X$ point due to strong electron-phonon coupling, suggesting a lattice instability with translational symmetry breaking. This is consistent with previous phonon dispersion calculation and supports a lattice distortion induced CDW picture\cite{6}. In the conventional CDW picture, the energy gap opens at the nested section of the Fermi surface, see Fig.4b, leading to a metal-insulator phase transition when all Fermi surfaces are gapped, such as transition metal dichalcogenides (TMDs)\cite{36}. However, the electronic reconstruction as revealed by ARPES can not be explained well by the Fermi surface nesting, but is in favor of a peculiar spectra weight redistribution\cite{4,27,28}. Here, based on our present NMR findings, we proposed that the $Ti$-$As$ bond order drives the electronic reconstruction. In Fig.4c, the calculated low-energy spin excitation spectrum in the tetragonal phase indicates that, besides a primary spin excitation around $Q=(0,0)$, there is also a substantial contribution from $Q\neq(0,0)$. This result suggests a possible spin-driven nematic fluctuation even without lattice instability. Therefore, although the phonon-induced lattice instability could lead to a conventional charge density-wave order, the nematic instability might overwhelm the charge density-wave instability and dominate the electronic reconstruction below the structural transition temperature. This is why the bond order instead of CDW order appears in this system. At the present stage, while it is difficult to disentangle the explicit role of lattice and nematic instabilities on the bond order, the bond order should be a consequence of the interplay of lattice and nematic instabilities.

Recently, a Hebel-Slichter coherent peak in 1/$(T_{1}T)$ has been observed in BaTi$_2$Sb$_2$O by NQR experiment\cite{24}, which supports a electron-phonon mediated Bardeen-Cooper-Schrieffer (BCS) superconductivity. Previous calculations indicated that the superconducting transition temperature $T_{c}$ due to electron-phonon coupling is very close to the experimental value of about 1$K$ in BaTi$_2$Sb$_2$O\cite{6}. Similar calculations also suggested phonon mediate superconductivity with \emph{T$_c$} about 8$K$ in {\cubbio}, which is however not observed in experiments\cite{6}. Based on the present findings, one possible explanation is that the bond order is a competing order against superconductivity. When the bond order is robust in {\cubbio} with higher \emph{T$_S$}, the expected BCS superconductivity is completely suppressed. This is quite similar to the relationship between nematic order and superconductivity in Fe-based superconductors\cite{29}. Further study on quantum melting of the bond order and its relationship with superconductivity could be a key to decode the mechanism of superconductivity in this family. On the other hand, the similar electronic reconstruction observed in BaTi$_2$As$_2$O has also been widely observed in many TMDs\cite{37,38}, which are ascribed to CDW order with strong electron-phonon coupling\cite{2}. Our present findings suggest that a possible orbital-relevant ordering could also emerge in these materials. It will be very interesting to revisit these CDW materials by searching orbital/bond order, which will promote the understanding on the underlying physics in TMDs.

\section*{ACKNOWLEDGEMENTS}
The authors are grateful for the stimulating discussions with Prof. Y. Li and Prof. Y. Liu and the help of Laue diffraction experiment from Dr. Y. J. Yan and Prof. D. L. Feng. This work is supported by the National Key R\&D Program of the MOST of China (Grant No. 2016YFA0300201, No. 2017YFA0303000, and No. 2016YFA0302300), the National Natural Science Foundation of China (Grants No. 11522434, U1532145, and 11674030), the Fundamental Research Funds for the Central Universities and the Chinese Academy of Sciences. The calculations used high performance computing clusters of Beijing Normal University in Zhuhai and the National Supercomputer Center in Guangzhou. T.W. and Z.P.Y acknowledge the Recruitment Program of Global Experts. T. W. acknowledges the CAS Hundred Talent Program.


\end{document}